\documentclass[superscriptaddress,twocolumn,showpacs,pre,floatfix]{revtex4-1}

\usepackage{amsmath}
\usepackage{color}
\usepackage{tabularx}
\usepackage[dvips]{graphicx}

\begin{document}

\title{The bimodal and Gaussian Ising Spin Glasses in dimension two revisited}

\author{P. H.~Lundow} \affiliation {Department of Mathematics and
  Mathematical Statistics, Ume{\aa} University, SE-901 87, Sweden}

\author{I. A.~Campbell} \affiliation{Laboratoire Charles Coulomb
  (L2C), UMR 5221 CNRS-Universit\'e de Montpellier, Montpellier,
  F-France.}

\begin{abstract}
A new analysis is given of numerical simulation data on the archetype
square lattice Ising Spin Glasses (ISG) with a bimodal ($\pm J$) and
Gaussian interaction distributions.  It is well established that the
ordering temperature of both models is zero. The Gaussian has a
non-degenerate ground state so exponent $\eta \equiv 0$ and it has a
continuous distribution of energy levels. For the bimodal model, above
a size-dependent cross-over temperature $T^{*}(L)$ there is a regime
of effectively continuous energy levels; below $T^{*}(L)$ there is a
distinct regime dominated by the highly degenerate ground state plus
an energy gap to the excited states. $T^{*}(L)$ tends to zero at very
large $L$ leaving only the effectively continuous regime in the
thermodynamic limit. We show that in this regime the critical exponent
$\eta$ is not zero, so the effectively continuous regime $2$D bimodal
ISG is not in the same universality class as the $2$D Gaussian
ISG. The simulation data on both models are analyzed using a scaling
variable $\tau = T^2/(1+T^2)$ suitable for zero temperature transition
ISGs, together with appropriate scaling expressions. Accurate
simulation estimates can be obtained for the temperature dependence of
the thermodynamic limit reduced susceptibility $\chi(\tau)$ and second
moment correlation length $\xi(\tau)$ over the entire range of
temperature from zero to infinity. The Gaussian critical exponent from
the simulations $\nu = 3.5(1)$ is in full agreement with the well
established value from the literature. The bimodal exponent from the
thermodynamic limit regime analysis is $\nu = 4.2(1)$, once again
different from the Gaussian value.
\end{abstract}


\maketitle

\section{Introduction}

The canonical Edwards-Anderson (EA) model Ising Spin Glasses (ISGs) in
dimension $d = 2$ have been the subject of very many numerical
studies. There is now consensus supported by analytic arguments that
the two archetype models, the ISGs on square lattices with near
neighbor interactions having distributions which are either Gaussian
or Bimodal ($\pm J$), have zero temperature transitions
\cite{hartmann:01,ohzeki:09}. For the Gaussian case, where the
interaction distribution is continuous and the ground state is unique,
there is now also general consensus concerning the low temperature
thermodynamic limit (ThL) behavior and exponents. In the bimodal case
there is an "effectively continuous energy level distribution" regime
coming down from high temperatures and ending with a crossover at a
size dependent temperature $T^{*}(L)$ to a ground state dominated
regime \cite{jorg:06}.  Interpretations differ considerably concerning
the critical exponents for the bimodal interaction case. We will give
a new analysis of accurate numerical Monte Carlo data up to size
$L=128$ on the bimodal system in the ThL and up to $L=64$ on the
Gaussian system. We use the temperature $T$ or the inverse temperature 
$\beta$ when it is convenient.

We first discuss discuss the specific heat using data from the
simulations and independent data down to low temperatures and large
sizes from Refs.~\cite{lukic:04} and \cite{thomas:11}. Then we analyse
the simulation data for other observables to obtain reliable and
accurate estimates for the critical exponents of the ThL regime, using
both the conventional scaling variable $T$ and a novel scaling
variable compatible with the generic scaling approach for ISGs
introduced in \cite{campbell:06}, adapted to a situation where $T_{c}
= 0$. We find that the Gaussian model and the bimodal model in the ThL
are not in the same universality class.

The $2$D Gaussian model is relatively clear-cut. Because $T_c = 0$ and
the interaction distribution is continuous, there is a unique ground
state (for each sample) and the low temperature excitation
distribution has no gap. The fact that the ground state is unique
necessarily implies that for all $L$, as $T \rightarrow 0$, $\xi(T,L)
\rightarrow \infty$ and the reduced susceptibility $\chi(T,L)
\rightarrow L^2$. With $T$ chosen as the critical scaling variable,
the standard thermodynamic limit low temperature critical expressions
are $\xi(T) \sim T^{-\nu}$ and $\chi(T) \sim T^{-\gamma} = T^{-2\nu}$
because the critical exponent $\eta$ is strictly zero. The critical
behavior of both the observables at low temperature is governed by the
single exponent $\nu$, which is related to the stiffness exponent
through $\theta = -1/\nu$. Accurate zero temperature domain wall
stiffness measurements to large sizes \cite{reiger:96, hartmann:01,
  carter:02, amoruso:03, houdayer:04, hartmann:02} show that $\theta = -0.285(2)$,
i.e. $\nu = 3.50(3)$.

In the $2$D bimodal case the situation is complicated by two
factors. First, the ground state is not unique but is massively
degenerate; the zero-temperature entropy per spin is $S_0 =
0.078(5)k_B$ \cite{hartmann:02,poulter:05,middleton:08}. Secondly, the
distribution of excited state energy levels is not continuous but
increases by steps of $4J$; in particular there is an energy gap $4J$
between the ground state and the first excited state. One can write
\cite{lukic:04} the "na\"ive" leading low temperature finite size
specific heat expression
\begin{equation}\label{Cvnaive}
  C_v(\beta,L) = \frac{16J^2\exp(S_1(L)-S_0(L))\exp(-4J/T)}{L^2T^2}
\end{equation}
where $S_1(L)$, $S_0(L)$ are the sample-averaged entropies of the first
excited state and the ground state. Setting $J=1$ a crossover
temperature can then be defined by
\cite{katzgraber:07,fisch:07,thomas:11}
\begin{equation}
T^{*}(L) = 4/(S_1(L)-S_0(L))
\end{equation}
which separates the critical behavior in the low temperature ground
state dominated regime (with $C_v(\beta,L) \sim \exp(-4/T)/T^2$) and a
$T > T^{*}(L)$ regime where the whole ensemble of higher energy states
dominate the thermodynamics \cite{jorg:06}.  An explicit
phenomenological expression for $T^{*}(L)$ derived from Eq.~5 of
Ref.~\cite{fisch:07}, which is consistent with the raw data points
\cite{lukic:04,fisch:07} for $S_1(L)-S_0(L)$ is
\begin{equation}\label{Tstar}
  \frac{4}{\exp(0.199\ln( 
    \frac{\ln(6.28L^2)}{2} + L^2(\ln(L^2)-1) 
     )+0.473)}
\end{equation}
A much simpler droplet-based expression from \cite{thomas:11} is
$T^{*}(L) \approx L^{-1/2}$.  $T^{*}(L)$ decreases with increasing $L$
because the degeneracy of the excited states increases faster with $L$
than that of the ground state. We will assume \cite{jorg:06} that in
the $T > T^{*}(L)$ ThL regime the data can be analysed in the same way
as if the energy level distribution were continuous.  With this
assumption the $T > T^{*}(L)$ regime will have "effectively
continuous" energy level distribution critical exponents with an
effective ordering temperature $T_c$ still zero. The ground state
dominated regime at $T < T^{*}(L)$ is a finite size effect which
disappears in the infinite $L$ limit.

A non-zero $\eta$ is to be expected {\it a priori} for a system with a
strong ground state degeneracy, unless each individual ground state is
isolated in phase space which is not the case \cite{hartmann:02} in
the bimodal ISG.  A droplet analysis of ground state measurements on
large sized samples \cite{hartmann:08} show that $\eta \approx 0.22$,
broadly consistent with a number of finite temperature simulation
estimates \cite{mcmillan:83,houdayer:01,katzgraber:05}.  However it
has been claimed that in the $T > T^{*}(L)$ regime the bimodal ISG can
be considered to be effectively in the same universality class as the
Gaussian ISG \cite{jorg:06}, meaning that the effective exponents are
again $\eta = 0$ and $\nu = 3.50(3)$. In view of the basic definition
of $\eta$ in terms of the short range limit of the spin-spin
correlation function $G(r,T) = G[r^{-\eta}\exp(r/\xi(T))]$, this claim
is rather surprising.

A major difficulty in establishing the limiting $[T > T^{*}(L), L
  \to\infty, T \to 0]$ behavior for ISGs in dimension $2$
\cite{katzgraber:07} consists in finding an appropriate and reliable
extrapolation procedure from simulation data necessarily restricted in
size and in temperature because of the need to achieve good thermal
equilibration at large sizes. This is a problem that we will address.

\section{Simulations}
The simulations were performed using the Houdayer cluster
method~\cite{houdayer:01} in combination with the exchange Monte
Carlo~\cite{hukushima:96} method. In the cluster step we first pick a
random site $i$ and compute its overlap $q_i=S^{A}_i S^{B}_i$, where
$S_i^{A}$ and $S_i^{B}$ denote the spin for two different replicas. We
then build an equal-$q$ cluster along the nearest neighbor
interactions and flip all cluster spins in both replicas. We used four
replicas which turned out to be remarkably efficacious.  On each
iteration the replicas are paired at random, then, for each pair, a
cluster update is performed, and the usual heat-bath spin update and
exchange.

For all systems we used $\beta_{\max}=3.0$.  The number of
temperatures were more than 250 for the smallest systems starting at
$\beta_{\min}=0.2$. With increasing system size the number of
temperatures was decreased and $\beta_{\min}$ increased. For the
largest system (bimodal $J_{ij}$ with $L=128$) $70$ temperatures were
used with $\beta_{\min}=1.2$. The exchange rate was always at least
$0.3$ for all systems and temperatures. The systems were deemed
equilibrated when the average $\langle q^2\rangle$ for the systems at
$\beta_{\max}$ appeared stable between runs. The number of
equilibration steps increased with system size, for the bimodal
$L=128$ this took about $600 000$ steps.  After equilibration, at
least $200 000$ measurements were made for each sample for all sizes,
taking place after every cluster-sweep-exchange step.

The usual observables were registered, the energy $E(\beta,L)$, the
correlation length $\xi(\beta,L)$, the spin overlap moments $\langle
|q| \rangle$, $\langle q^2\rangle$, $\langle |q|^3\rangle$, $\langle
q^4\rangle$. Correlations $\langle E(\beta,L), U(\beta,L)\rangle$
between the energy and some observables $U(\beta,L)$ were also
registered. Thermodynamic derivatives could then be evaluated through
the usual $\partial U(\beta,L)/\partial \beta = \langle U(\beta,L),
E(\beta,L)\rangle-\langle U(\beta,L) \rangle\langle
E(\beta,L)\rangle$. Error estimates of observables and derivatives
were done with the bootstrap method.

Sizes studied were $L=4$, $6$, $8$, $12$, $16$, $24$, $32$, $48$,
$64$, $96$, $128$ for bimodal interactions, and up to $L=64$ for the
Gaussian interactions, with $2^{13}=8192$ samples
($J_{ij}$-interactions) for all sizes.

\section{Specific heat}

The size dependence of the ground state energy per spin $e(0,L)$ for
the 2D Gaussian ISG has been shown \cite{campbell:04,middleton:08} to
follow the simple critical finite size scaling rule
\begin{equation}
e(0,L) - e(0,\infty) \sim L^{2-\theta}
\end{equation}
with a $\theta$ consistent with the estimate from ground state domain
wall stiffness measurements \cite{hartmann:01}.  Standard scaling
arguments \cite{lukic:04} would suggest that the low temperature
specific heat should behave as
\begin{equation}
  C_v(\beta,L) \sim \beta^{-2\nu} \approx \beta^{-7}
\end{equation}
but because of the continuous interaction distribution, in addition to
critical excitations there are always single spin excitations. These
lead to a term $C_v(\beta,L) \approx T$ which dominates the Gaussian low
temperature specific heat as noted by Ref.~\cite{lukic:04}.
Specific heat data for the bimodal model were calculated through the
present simulations; data extending to a much lower temperature range
and larger sizes have already been measured using the sophisticated
Pfaffian arithmetic technique by Lukic {\it et al}~\cite{lukic:04} and
by Thomas {\it et al}~\cite{thomas:11}, and we are very grateful to be
able to quote these results {\it in extenso}.

\begin{figure}
  \includegraphics[width=3.0in]{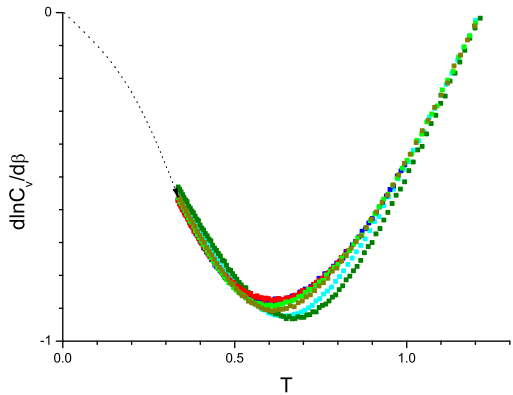}
  \caption{(Color on line) Gaussian 2D ISG. Logarithmic derivative of
    the specific heat $\partial\ln C_{v}(T,L)/\partial\beta$
    against $T$. Sizes $L = 64$, $48$, $32$, $24$, $16$, $12$ top to
    bottom in the dip. Curve : extrapolation.} \protect\label{fig1}
\end{figure}

\begin{figure}
  \includegraphics[width=3.0in]{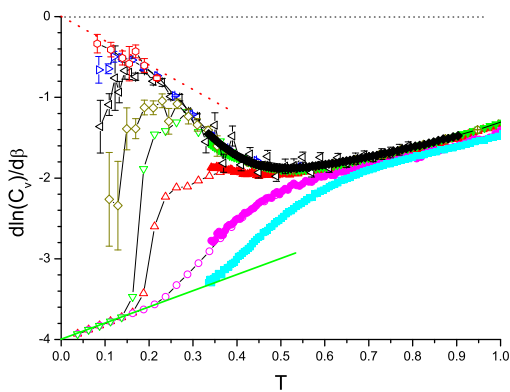}
  \caption{(Color on line) Bimodal 2D ISG. Logarithmic derivative of
    the specific heat $\partial\ln C_{v}(T,L)/\partial\beta$
    against $T$.  Full points : simulation data $L = 96$, $48$, $24$,
    $12$, $8$ (black, green, red, pink, cyan) top to bottom. Open
    points : Pfaffian data ; red polygons $L=512$, blue right
    triangles $L =256$, black left triangles $L=128$, brown diamonds
    $L=64$ (all data from Ref.~\cite{thomas:11}), green down triangles
    $L=50$, red up triangles $L=24$, pink circles $L=12$, all data fom
    Ref.~\cite{lukic:04}. Dashed diagonal red line $y= -3x$, green
    diagonal line $y=-4+2x$.}\protect\label{fig2}
\end{figure}

The data for the two models are shown, see Fig.~\ref{fig1} and
Fig.~\ref{fig2}, in the form of plots of the derivative
$y=\partial\ln(C_v(\beta,L))/\partial\beta$ against $x=T$.  This
non-conventional form of plot happens to be particularly
instructive. A low temperature limit $C_v(\beta,L) \sim T^{x}$ appears
as a straight line through the origin with slope $-x$, while a low
temperature limit of the "na\"{i}ve" bimodal ground state dominated
form Eq.~\eqref{Cvnaive} appears as a straight line with intercept
$-4$ and slope $+2$.

The Gaussian data are almost independent of $L$ for the whole
temperature range. Physically this occurs because the specific heat in
ISGs is predominately a near neighbor effect. The curve tends to a
slope $\partial y/\partial x \sim -1$ corresponding to $C_{v} \sim
T^{1}$ in the low $T$ limit, in agreement with the conclusion of
Ref.~\cite{jorg:06}.

For the bimodal model there is first a high temperature and/or high
$L$ envelope curve corresponding to the effectively continuous $T >
T^{*}(L)$ regime. In this regime finite size effects are very weak :
the specific heat is almost independent of $L$ as in the Gaussian. The
curves for the two models are of similar form but are not
identical. In the large $L$, low $T$ limit of this envelope curve, the
bimodal data as shown in Fig.~\ref{fig2} indicate $C_v\sim T^{3}$ in
agreement with the conclusions drawn in Ref.~\cite{thomas:11} based on
droplet excitation arguments.

For each $L$ the data curve peels off the large $L$ envelope curve
below an $L$-dependent temperature which can be identified with the
start of the effectively continuous to ground state dominated regime
crossover centered at $T^{*}(L)$.
Finally for each $L$ in the low temperature range $T \ll T^{*}(L)$ the
specific heat links up to the "na\"ive" limit of
Eq.~\eqref{Cvnaive}. (It should be noted that because of the
logarithmic derivative, temperature independent $L$-dependent factors
do not show up in this plot). The crossover can be seen to be gentle
for small $L$, becoming sharp for large $L$.  Defining $T^{*}(L)$ as
the location of the maximum positive slope on this plot, the crossover
temperatures can be clearly identified and are consistent with
$T^{*}(L)L^{1/2} = 1.1(1)$.

An anomalous limit of the form
\begin{equation}
  C_v(\beta) \sim \beta^2\exp(-2\beta)
\end{equation}
which has been proposed by some authors \cite{lukic:04,wang:05}
following Ref.~\cite{wang:88} is inconsistent with the data in
Fig.~\ref{fig2} for all $L$ and $T$ (see also
\cite{saul:93,poulter:08}). An intermediate $L$ regime where
$C_v(\beta,L) \sim T^{5.25}$ as proposed in Ref.~\cite{fisch:07} or
$C_v(\beta,L) \sim T^{4.2}$ as proposed in Ref.~\cite{katzgraber:07}
appear to be valid only for a limited range of $T$ and $L$.

\section{The exponent $\eta$}

For ISGs with non-zero critical temperatures finite size scaling
analyses at and close to the critical temperature are used to estimate
critical exponents. For the $2$D bimodal ISG, because of the crossover
to the ground state dominated regime this approach is ruled out and
the critical exponents must be estimated using the entirely different
strategy of ThL measurements.

The standard Renormalization Group theory (RGT) scaling variable for
models with non-zero ordering temperatures is
$t=(T-T_{c})/T_{c}$. This obviously cannot be used when $T_{c} = 0$;
by convention the scaling variable used in the literature for $2$D
ISGs is the un-normalized temperature $T$. This is only a convention;
it is perfectly legitimate to use other conventions. Thus, when
considering the canonical $1$D Ising ferromagnet, Baxter
\cite{baxter:82} remarks "When $T_{c} = 0$ it is more sensible to
replace $t=(T-T_c)/T_c$ by $\tau = \exp(-2\beta)$". (In fact for the
particular $1$D model scaling without corrections over the entire
temperature range follows if a related scaling variable $\tau=
1-\tanh(\beta)$ is chosen \cite{katzgraber:08,campbell:11}). Below we
will introduce another scaling variable appropriate for ISGs with
$T_c=0$, but for the moment we follow this traditional $t=T$ $2$D ISG
convention. The critical exponents are defined through the leading ThL
expressions for the reduced susceptibility and the second moment
correlation length within this convention : $\chi(T)
=C_{\chi}T^{-(2-\eta)\nu}$ and $\xi(T) =C_{\xi}T^{-\nu}$ in the limit
$T \to 0, L \to \infty$.  For all data which fulfil the condition
(either in the bimodal and Gaussian models) $L > K\xi(T,L)$ with $K
~\approx 6$, observables such as $\chi(T,L)$ and $\xi(T,L)$ depend on
$T$ but not on $L$, and so correspond to the ThL infinite size values
$\chi(T)$ and $\xi(T)$. The ThL condition defines implicitly a
crossover temperature $T_{\xi}(L)$. It turns out that in the bimodal
$2$D ISG the ThL limit temperature $T_{\xi}(L)$ is always higher than
the corresponding crossover temperature to the ground state dominated
regime $T^{*}(L)$ defined above, so the ThL data are always well in
the effectively continuous regime. The ThL data extrapolation to $T =
0$ corresponds to estimates for the critical exponents in the
successive limits $[L \to \infty, T \to 0]$ so in the effectively
continuous energy level regime, to be distinguished from the exponents
defined taking the successive limits $[T \to 0, L \to \infty]$ which
would correspond to the "finite size" ground state dominated regime.

There have been many previous studies having the aim of estimating the
critical exponents and in particular $\eta$ for the bimodal
model. McMillan already in 1983 estimated $\eta=0.28(4)$ from $G(r)$
correlation data on one $L=96$ sample well in the effectively
continuous regime \cite{mcmillan:83}. Katzgraber and Lee
\cite{katzgraber:05} estimated $\eta=0.138(5)$ from $\chi(T,L)$
data. J\"org {\it et al} \cite{jorg:06} show a plot of $\ln\chi(T)$
against $\ln\xi(T)$ after an extrapolation to infinite $L$ using the
technique of Ref.~\cite{caracciolo:95}. They state "fits of this curve
lead to values of $\eta$ that are very small, between $0$ and $0.1$,
strongly suggestive of $\eta = 0$". However, this type of
extrapolation to infinite $L$ is delicate, particularly in the bimodal
$2$D case.


In addition, the data displayed by \cite{jorg:06} on a
$\ln\chi(T)-\ln\xi(T)$ plot extending over five decades on the $y$
axis are hard to fit with precision. Katzgraber {\it et al}
\cite{katzgraber:07} show a plot of
$\partial\ln\chi(T,L)/\partial\ln\xi(T,L)$ which in principle is
equivalent to the Ref.~\cite{jorg:06} plot but which provides a
display much more sensitive to the value of $\eta$; they state
cautiously "for all system sizes and temperatures studied
$\eta_{\mathrm{eff}}$ is always greater than $0.2$, although an
extrapolation to $\eta = 0$ cannot be ruled out", so that the
possibility of the bimodal and Gaussian ISGs being in the same
universality class "cannot be reliably proven". In
Refs.~\cite{toldin:10,toldin:11} it is claimed that the Gaussian and
bimodal models are in the same universality class, which is surprising
as "the data are not sufficiently precise to provide a precise
determination of $\eta$, being consistent with a small value $\eta
\leq 0.2$, including $\eta = 0$".

All the estimates quoted so far can be considered to concern the
effectively continuous regime. At zero or low temperatures, so in the
ground state dominated regime, different sophisticated algorithms lead
to the estimates $\eta=0.14(1)$ \cite{poulter:05}, and to $\eta=0.22$
\cite{hartmann:08}.

\begin{figure}
  \includegraphics[width=3.0in]{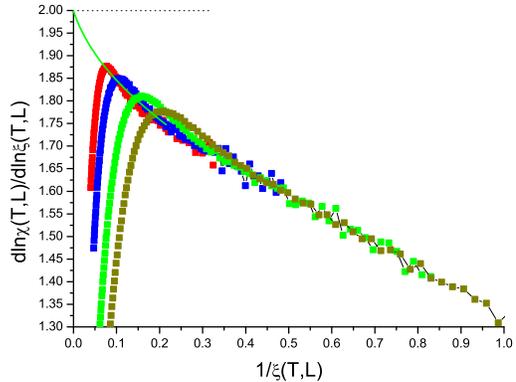}
  \caption{(Color on line) Gaussian 2D ISG. Derivative
    $\partial\ln\chi(T,L)/\partial\ln\xi(T,L)$ against $1/\xi(T,L)$
    for $L = 64$, $48$, $32$, $24$ left to right. In this and all
    following figures both Gaussian and bimodal, the color coding is :
    black, pink, red, blue, green, brown, cyan, olive for $L = 128$,
    $96$, $64$, $48$, $32$, $24$, $16$, $12$.}\protect\label{fig3}
\end{figure}

\begin{figure}
  \includegraphics[width=3.0in]{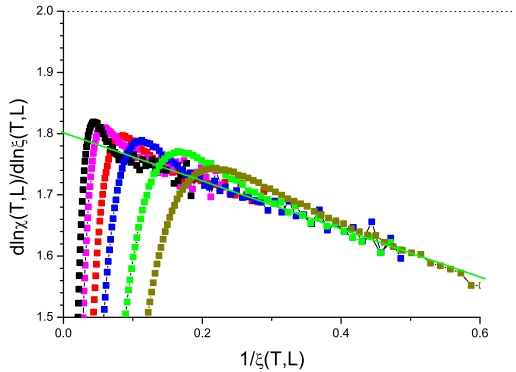}
  \caption{(Color on line) Bimodal 2D ISG. Derivative
    $\partial\ln\chi(T,L)/\partial\ln\xi(T,L)$ against $1/\xi(T,L)$
    for $L = 128$, $96$, $64$, $48$, $32$, $24$ left to right. Same
    color coding as in Fig.~\ref{fig3}.}\protect\label{fig4}
\end{figure}

In Fig.~\ref{fig3} and Fig.~\ref{fig4}, we show plots of $y =
\partial\ln\chi(T,L)/\partial\ln\xi(T,L)$ against $x =1/\xi(T,L)$ for
the Gaussian and bimodal models. These are {\it raw} data points
having the high statistical precision of the present measurements.
With the conventional definition of the critical exponents through
$\chi(T,L) \sim T^{−(2−\eta)\nu}$ and $\xi(T,L) \sim T^{−\nu}$ in the
ThL regime low-$T$ limit, the limiting slope $\partial y/\partial x$
at criticality as $x\to 0$ is by definition equal to $2-\eta$. For the
Gaussian model the observed tendency of the slope is consistent with
the limit of $\eta=0$ which must be the case for this nondegenerate
ground state model. For the bimodal model the observed $y(x)$ in the
ThL regime is not tending to $2$ but to a constant limit of
$1.80(2)$. Slight overshoots for each $L$ in both systems can be
ascribed to $\chi(T,L)$ and $\xi(T,L)$ not reaching the ThL condition
at quite the same temperature. As stated above, very similar
observations were made in Ref.~\cite{katzgraber:07} for the bimodal
model.  The present results thus confirm unambiguously that for the
bimodal ISG in the effectively continuous regime $\eta$ is not zero
but is $\approx 0.20$. Thus the bimodal ISG in the effectively
continuous ThL regime and the Gaussian ISG are not in the same
universality class.


It has been shown that in dimension $4$ also, Gaussian and bimodal
ISGs are not in the same universality class \cite{lundow:15}, so the
breakdown of universality in ISGs appears to be general.

\section{Scaling and zero temperature ordering}

Estimating the exponents $\nu$ or $\gamma=(2-\eta)\nu$ is more
difficult than for the exponent $\eta$.  As we have noted above, the
standard RGT convention for models with finite temperature ordering is
to use the scaling variable $t = (T/T_c)-1$, which obviously cannot be
applied to models with $T_c =0$, and for $2$D ISGs the preferred
convention in the literature has been to use the un-normalized scaling
variable $t=T$. In practice this is inefficient as the extrapolations
towards the $T = 0$ limit in order to estimate the values for the
critical exponents are very ambiguous. For instance, when presenting
$T$-scaled susceptibility data for sizes up to $L=128$ Katzgraber {\it
  et al} \cite{katzgraber:07} state "the [susceptibility] data for the
bimodal case can be extrapolated to any arbitrary value including
$1/\gamma_{\mathrm{eff}} = 0$".

We will introduce a novel scaling variable suitable for the $2$D ISGs,
applying the same principles as for ISGs at higher dimensions
\cite{campbell:06}, adapted to $T=0$ ordering :

-- For spin glasses the relevant interaction strength parameter is not
$J$ but is $\langle J^2 \rangle$, so the natural dimensionless
parameter is $\langle J^2 \rangle\beta^2$ (or alternatively
$\tanh^2(J\beta)$ for bimodal ISGs). With the standard normalisation
$\langle J^2 \rangle\ =1$ the natural inverse "temperature" in ISGs is
$\beta^2$, not $\beta$. This was recognized immediately after the
Edwards-Anderson model was introduced, in high temperature series
expansion (HTSE) analyses for ISGs including $2$D models
\cite{singh:87,klein:91,daboul:04}, but has since been overlooked in
most simulation analyses.

-- It is convenient to choose a scaling variable $\tau$ defined in
such a way that $\tau=0$ at criticality and $\tau=1$ at infinite
temperature. With an ISG ordering at a finite inverse temperature
$\beta_{c}$, $\tau(\beta) = 1-\beta^2/\beta_{c}^2$ is an appropriate
choice \cite{daboul:04,campbell:06}. When $\beta_{c} = \infty$ as in
the $2$D ISG case, $\tau_{t}(\beta) =
1-(\tanh(\beta)/\tanh(\beta_{c}))^2 = 1-\tanh(\beta)^2$ has been used
\cite{singh:87}, but here we will prefer $\tau_{b}(\beta) =
1/(1+\beta^2)$ as it turns out to be efficient and the limits are easy
to relate to those of the $T$ scaling convention. With non-zero $T_c$
the effective exponents at criticality do not depend on the choice of
scaling variable; this is not the case when $T_{c}=0$, but a simple
dictionary is given below relating the limiting derivatives for
$\tau_{b}$ scaling to the exponents for the conventional $T$ scaling.

-- The ThL HTSE Darboux \cite{darboux:78} format for observables $Q(x)$ is
\begin{equation}
Q(x) = 1 + a_{1}x + a_{2}x^2 + a_{3}x^3 + \cdots
\end{equation}
with $x = \beta^2$ in ISGs \cite{daboul:04}. The HTSE ISG
susceptibility $\chi(\beta^2)$ is naturally in this format, so for ISG
models with $T_{c} > 0$ the ThL susceptibility can be scaled in the
Wegner \cite{wegner:72} form
\begin{equation}
\chi(\beta^2) = C_{\chi}\tau(\beta^2)^{-(2-\eta)\nu}F[1+a_{\chi}\tau(\beta^2)^{\theta}+\cdots]
\end{equation}
Because the correlation function second moment $\mu_{2}$ HTSE is of
the form (see Ref.~\cite{butera:02} for the Ising ferromagnet)
\begin{equation}
\mu_{2}(x) = x + a_{1}x^2 + a_{2}x^3 +\cdots
\end{equation}
and the second moment correlation length is defined through
$\xi(x)^2=\mu_{2}/(z\chi(x))$ with $z$ the number of nearest
neighbors, for consistency the appropriate correlation length variable
for ISG scaling is $\xi(x)/\beta$ rather than $\xi(x)$ (whether $T_{c}$
is zero or not). This point has been spelt out in
Ref.~\cite{campbell:06}.

Examples of applications of the scaling rules outlined here to other
specific models (both ferromagnets and ISGs) have been given
elsewhere.  A general discussion of ferromagnets and spin glasses is
given in Ref.~\cite{campbell:06}, analyses of $3$D Ising, XY and
Heisenberg ferromagnets in Ref.~\cite{campbell:07}, the $2$D Ising
ferromagnet is analysed in Ref.~\cite{campbell:08}, $3$D Ising
ferromagnets in \cite{campbell:11,lundow:11}, high dimension Ising
ferromagnets in Ref.~\cite{berche:08}, and the $2$D Villain fully
frustrated model in Ref.~\cite{katzgraber:08}.

The scaling of the Binder cumulant 
\begin{equation}\label{gdef}
  g(\beta^2) = \frac{1}{2} \left(3-\frac{\lbrack\langle
  q^4\rangle\rbrack}{\lbrack\langle q^2\rangle\rbrack^2}\right)
\end{equation}
is discussed in the Appendix.  The $2$D simulation data analysis and
the extrapolations below are based on the derivatives $\partial\ln
Q(\tau_{b},L)/\partial\ln\tau_{b}$ in the ThL regime where these
derivatives are independent of $L$ and so equal to the infinite size
derivatives. An advantage of the $2$D models is that in contrast to
$\tau(\beta^2)$ for the models with non-zero $T_{c}$, for the $2$D
ISGs with $T_{c} \equiv 0$ there is no uncertainty in the definition
of $\tau_{b}(\beta^2)$ related to an uncertainty in the value of the
ordering temperature.

Once the $\tau_{b} \rightarrow 0$ limits for the various derivatives
have been estimated by extrapolation of the ThL data for finite $L$,
there is a simple dictionary for translating into terms of the
conventional $T$ scaling critical exponents $\nu$ and $\eta$ defined
above :

\begin{subequations}\label{derivs}
  \begin{eqnarray}
    -\frac{\partial\ln\chi(\tau_{b})}{\partial\ln\tau_{b}}&\to&  \frac{\nu(2-\eta)}{2}\label{deriv1}\\
    -\frac{\partial\ln(T\xi(\tau_{b}))}{\partial\ln\tau_{b}}&\to&  \frac{(\nu-1)}{2}\label{deriv2}\\
    \frac{\partial\ln\chi(\tau_{b})}{\partial\ln(T\xi(\tau_{b}))}&\to& \frac{\nu(2-\eta)}{\nu-1}\label{deriv3}\\
    -\frac{\partial\ln g(\tau_{b})}{\partial\ln\tau_{b}}&\to& \nu\label{deriv4}
  \end{eqnarray}
\end{subequations}

\section{Analyses with the scaling variable $\tau_{b}$}

\begin{figure}
  \includegraphics[width=3.0in]{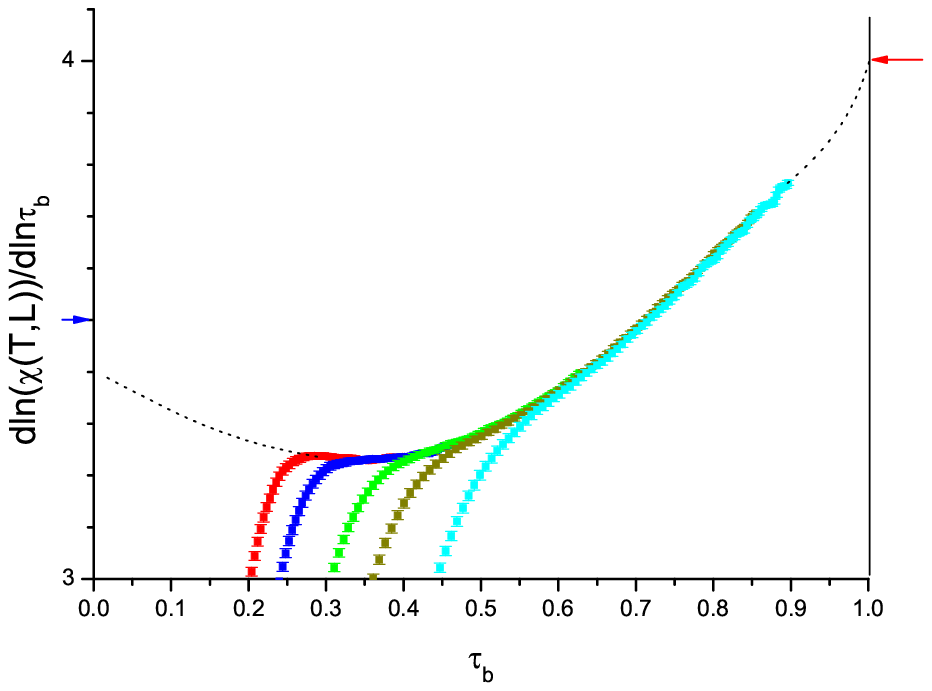}
  \caption{(Color on line) Gaussian 2D ISG. The derivative
    $\partial\ln\chi(T,L)/\partial\ln\tau_{b}$ against $\tau_{b}$.
    Sizes $L = 64$, $48$, $32$, $24$, $16$ left to right. Same color
    coding as in Fig.~\ref{fig3}. Dashed line: extrapolation. Red
    arrow : exact infinite temperature value. Blue arrow : Gaussian
    critical value.}\protect\label{fig5}
\end{figure}

\begin{figure}
  \includegraphics[width=3.0in]{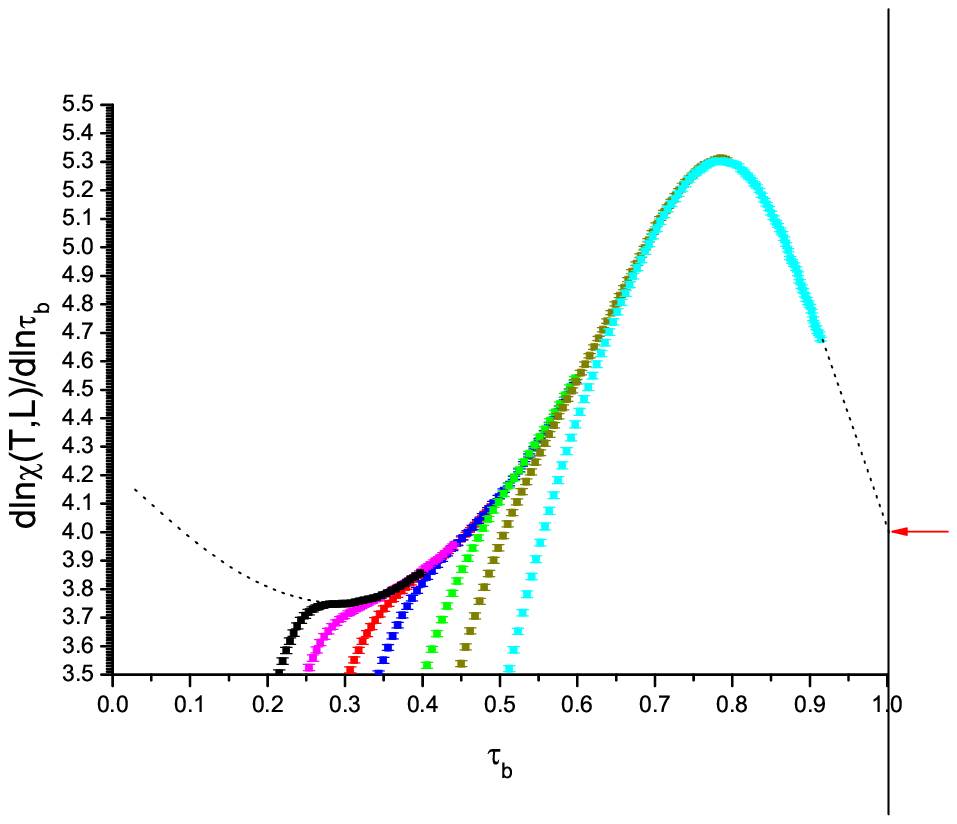}
  \caption{(Color on line) Bimodal 2D ISG. The derivative
    $\partial\ln\chi(T,L)/\partial\ln\tau_{b}$ against $\tau_{b}$.
    Sizes $L = 128$, $96$, $64$, $48$, $32$, $24$, $16$ left to
    right. Same color coding as in Fig.~\ref{fig3}. Dashed line:
    extrapolation. Red arrow : exact infinite temperature value.
  }\protect\label{fig6}
\end{figure}

\begin{figure}
  \includegraphics[width=3.0in]{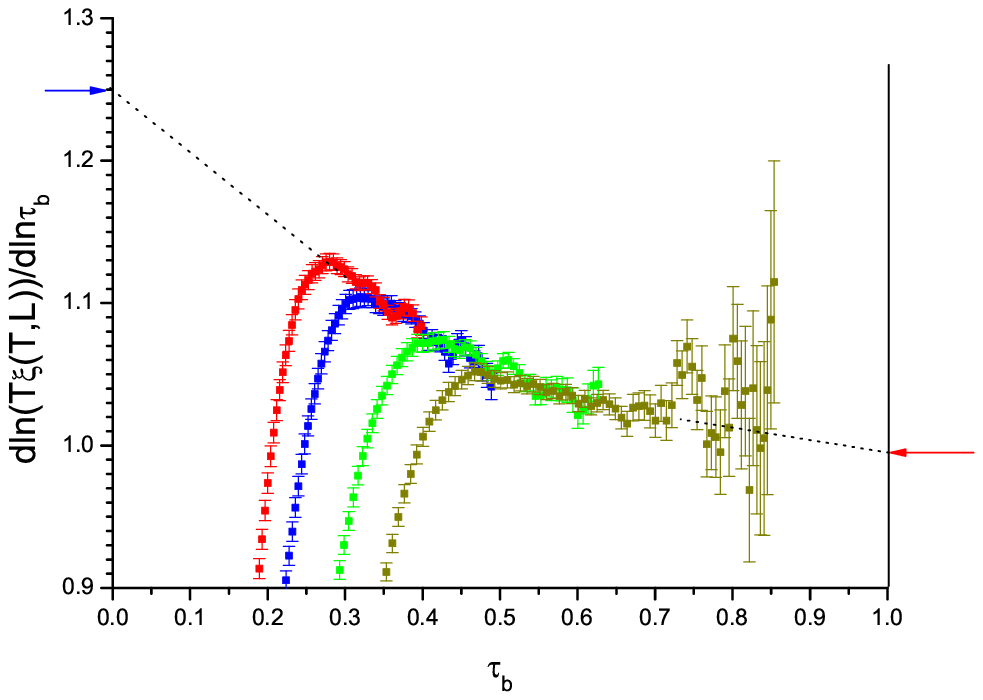}
  \caption{(Color on line) Gaussian 2D ISG. The derivative
    $\partial\ln(T\xi(T,L))/\partial\ln\tau_{b}$ against $\tau_{b}$.
    Sizes $L = 64$, $48$, $32$, $24$ left to right. Same color
    coding as in Fig.~\ref{fig3}. Dashed line: extrapolation. Red
    arrow : exact infinite temperature value. Blue arrow : Gaussian
    critical value.}\protect\label{fig7}
\end{figure}

\begin{figure}
  \includegraphics[width=3.0in]{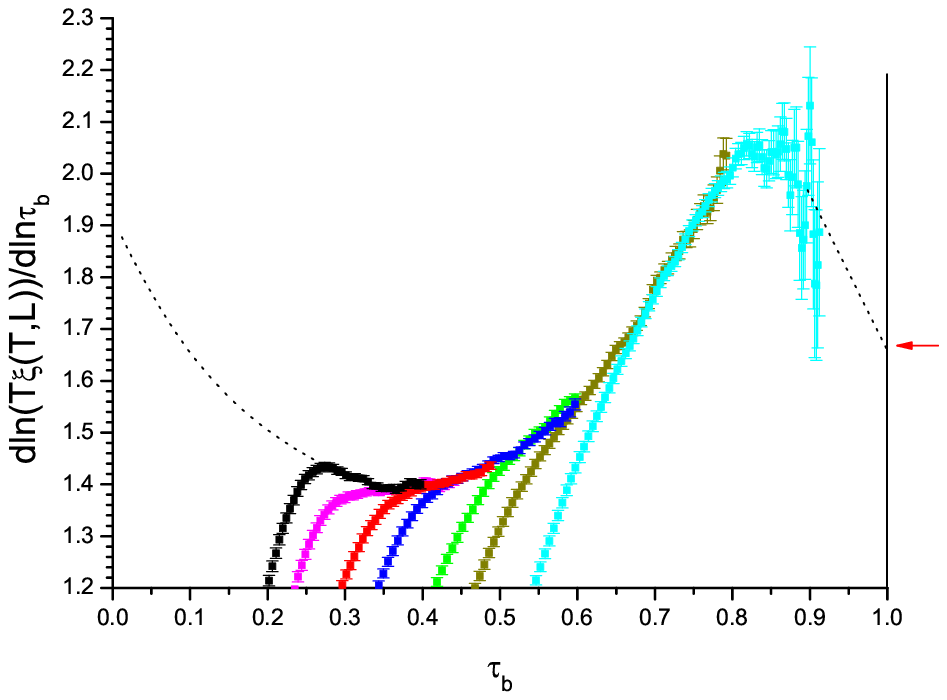}
  \caption{(Color on line) Bimodal 2D ISG. The derivative
    $\partial\ln(T\xi(T,L))/\partial\ln\tau_{b}$ against
    $\tau_{b}$. Sizes $L = 128$, $96$, $64$, $48$, $32$, $24$, $16$
    left to right. Same color coding as in Fig.~\ref{fig3}. Dashed
    line: extrapolation. Red arrow : exact infinite temperature value.
  }\protect\label{fig8}
\end{figure}

\begin{figure}
  \includegraphics[width=3.0in]{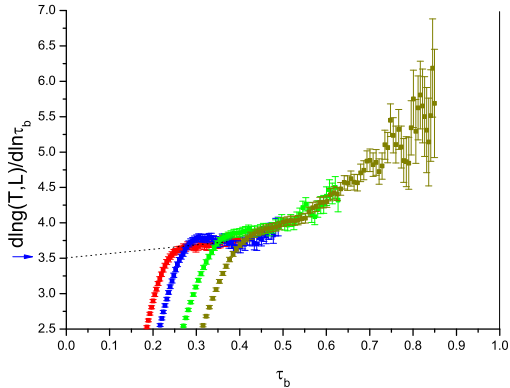}
  \caption{(Color on line) Gaussian 2D ISG. The derivative
    $\partial\ln g(T,L)/\partial\ln\tau_{b}$ against $\tau_{b}$, where
    $g(T,L)$ is the Binder cumulant. $L= 64$, $48$, $32$, $24$ left to
    right. Same color coding as in Fig.~\ref{fig3}. Line:
    extrapolation. Blue arrow : Gaussian critical value.
  }\protect\label{fig9}
\end{figure}

\begin{figure}
  \includegraphics[width=3.0in]{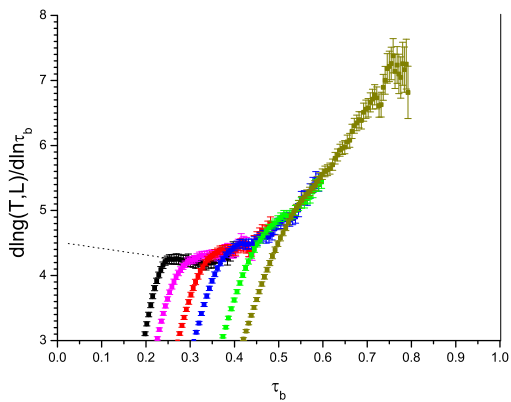}
  \caption{(Color on line) Bimodal 2D ISG. The derivative $\partial\ln
    g(T,L)/\partial\ln\tau_{b}$ against $\tau_{b}$, where $g(T,L)$ is
    the Binder cumulant. Sizes $L= 128$, $96$, $64$, $48$, $32$, $24$
    left to right. Same color coding as in Fig.~\ref{fig3}. Line:
    extrapolation.  }\protect\label{fig10}
\end{figure}

\begin{figure}
  \includegraphics[width=3.0in]{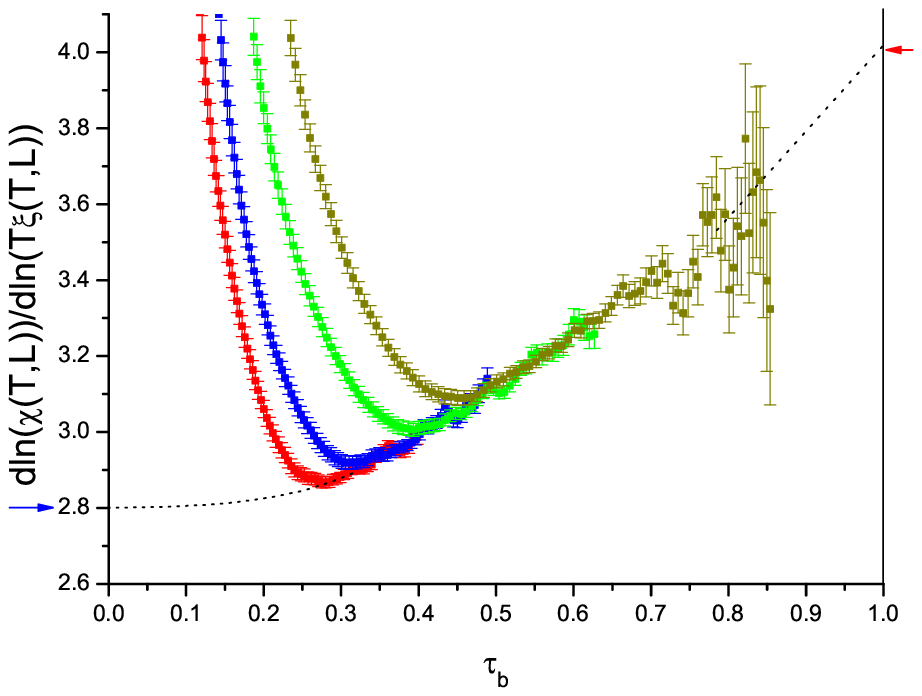}
  \caption{(Color on line) Gaussian 2D ISG. The derivative
    $\partial\ln\chi(T,L)/\partial\ln(T\xi(T,L))$ against
    $\tau_{b}$. Sizes $L = 64$, $48$, $32$, $24$ left to right. Same
    color coding as in Fig.~\ref{fig3}. Dashed line:
    extrapolation. Red arrow : exact infinite temperature value.  Blue
    arrow : Gaussian critical value.}\protect\label{fig11}
\end{figure}

\begin{figure}
  \includegraphics[width=3.0in]{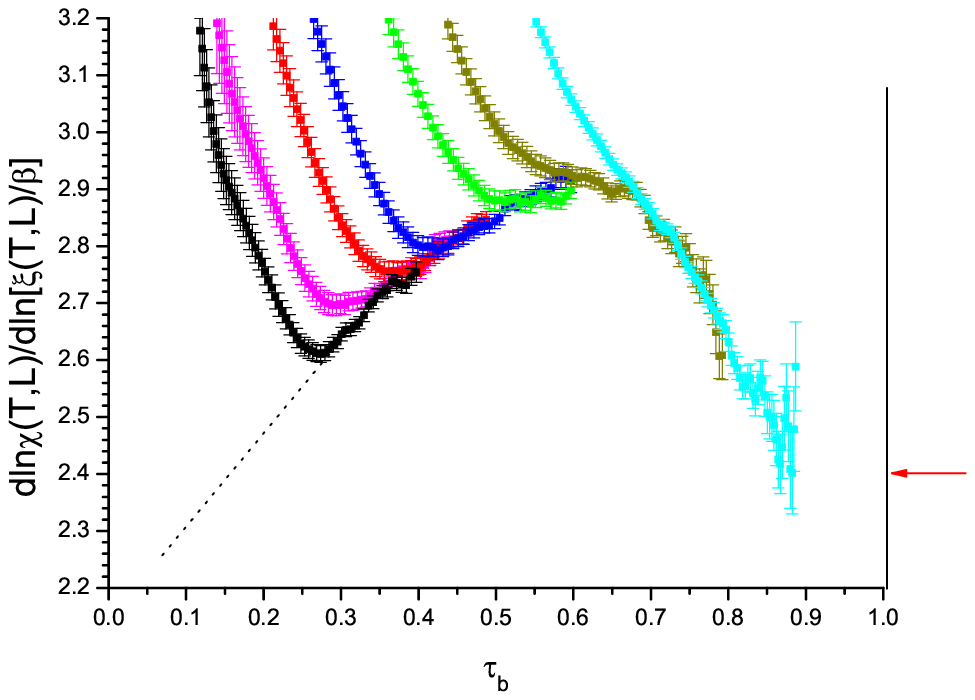}
  \caption{(Color on line) Bimodal 2D ISG. The derivative
    $\partial\ln\chi(T,L)/\partial\ln(T\xi(T,L))$ against
    $\tau_{b}$. Sizes $L = 128$, $96$, $64$, $48$, $32$, $24$, $16$
    left to right. Same color coding as in Fig.~\ref{fig3}. Dashed
    line: extrapolation. Red arrow : exact infinite temperature value.
  }\protect\label{fig12}
\end{figure}

The four derivatives of Eq.~\eqref{derivs}
are shown in
Figs.~\ref{fig5} to \ref{fig12}. In contrast to the derivatives in
which $T$ is used as the scaling variable, each derivative can be
extrapolated in a fairly unambiguous manner to criticality, and always
has an exact finite value at infinite temperature $\tau_{b} = 1$ .

The exact infinite temperature limits from the general high
temperature scaling expansion expressions \cite{daboul:04} applied to
scaling with $\tau_{b}$ are (when $\tau_b\to 1$):
\begin{subequations}\label{exact}
\begin{eqnarray}
  -\frac{\partial\ln\chi(\tau_{b})}{\partial\ln\tau_{b}} &=&4 
  \,\textrm{(Gauss.)}, \ldots =4\,\textrm{(bimodal)}\label{exact1}\\ 
  \frac{\partial\ln(T\xi(\tau_{b}))}{\partial\ln\tau_{b}} &=&1 
  \,\textrm{(Gauss.)}, \ldots =\frac{5}{3}\,\textrm{(bimodal)}\label{exact2}\\ 
  \frac{\partial\ln\chi(\tau_{b})}{\partial\ln(T\xi(\tau_{b}))} &=&4 
  \,\textrm{(Gauss.)}, \ldots =\frac{12}{5} \,\textrm{(bimodal)}\label{exact3}
\end{eqnarray}
\end{subequations}

The extrapolation method is outlined in Appendix II. With $\eta=0$ and
assuming $\nu=3.48(5)$ \cite{hartmann:02}, the predicted Gaussian
critical limits (when $\tau_b\to 0$) for the derivatives are
\begin{subequations}\label{gderivs}
  \begin{eqnarray}
    -\frac{\partial\ln\chi(\tau_{b})}{\partial\ln\tau_{b}}&\to& 3.48(5)\label{gderiv1}\\
    -\frac{\partial\ln(T\xi(\tau_{b}))}{\partial\ln\tau_{b}}&\to&  1.24(3)\label{gderiv2}\\
    \frac{\partial\ln\chi(\tau_{b})}{\partial\ln(T\xi(\tau_{b}))}&\to& 2.81(1)\label{gderiv3}\\
    -\frac{\partial\ln g(\tau_{b})}{\partial\ln\tau_{b}}&\to& 3.5(3)\label{gderiv4}
  \end{eqnarray}
\end{subequations}
From the fitted ThL data extrapolations (see Appendix II) the
estimated Gaussian critical limit values are $3.40(10)$, $1.25(5)$,
$2.90(10)$, $3.6(1)$ respectively.  These values are fully consistent
with the list above, validating the approach and the extrapolation
procedure that we have used.  For the bimodal data, the extrapolated
ThL limits (when $\tau_b\to 0$) from the figures (see Appendix II)
give estimates
\begin{subequations}\label{jderivs}
  \begin{eqnarray}
    -\frac{\partial\ln\chi(\tau_{b})}{\partial\ln\tau_{b}}&\to&  4.3(1)\label{jderiv1}\\
    -\frac{\partial\ln(T\xi(\tau_{b}))}{\partial\ln\tau_{b}}&\to&  1.9(1)\label{jderiv2}\\
    \frac{\partial\ln\chi(\tau_{b})}{\partial\ln(T\xi(\tau_{b}))}&\to& 2.15(10)\label{jderiv3}\\
    -\frac{\partial\ln g(\tau_{b})}{\partial\ln\tau_{b}}& \to& 4.7(1)\label{jderiv4}
  \end{eqnarray}
\end{subequations}
These are all significantly different from the Gaussian values,
confirming non-universality.  When translated into the $T$ scaling
convention,
the 2D bimodal critical exponents from these measurements are $\eta =
0.20(2)$ and $\nu = 4.8(2)$ (so $\gamma = (2- \eta)\nu = 8.6(4))$.
Not only is the ThL bimodal exponent $\eta$ different from
the Gaussian value but the $\nu$ value is different also.

\section{Conclusion}

Simulation data are presented for the Gaussian and bimodal interaction
distribution Ising spin glasses in dimension two, which are known to
order only at zero temperature. In order to facilitate extrapolations
to zero temperature, a temperature scaling variable $\tau_{b}=
T^2/(1+T^2)$ is introduced in addition to the traditional $2$D ISG
scaling variable $t=T$.  The Gaussian simulation data are completely
consistent with the well established critical behavior for this model,
with exponents $\eta \equiv 0$ and $\nu = 3.48(5)$ \cite{hartmann:02}.

The bimodal specific heat simulation data supplemented by data from
Lukic {\it et al} \cite{lukic:04} and from Thomas {\it et al}
\cite{thomas:11} show clear crossovers from an effectively continuous
energy level thermodynamic limit regime to a finite size ground state
dominated regime at size dependent temperatures $T^{*}(L) \approx
1.1/L^{1/2}$ (see Ref.~\cite{thomas:11}). 

The extrapolated ThL simulation results tend to critical limits which
correspond consistently to $\eta = 0.20(2)$ and $\nu = 4.8(2)$,
clearly different from the Gaussian values. This demonstrates that the
standard universality rules do not hold for $2$D ISG models. In
dimension $4$ also bimodal and Gaussian ISGs have been shown not to be
in the same universality class either \cite{lundow:15}, strongly
suggesting a lack of universality for ISGs in each dimension
(presumably up to the upper critical dimension).

\begin{figure}
  \includegraphics[width=3.0in]{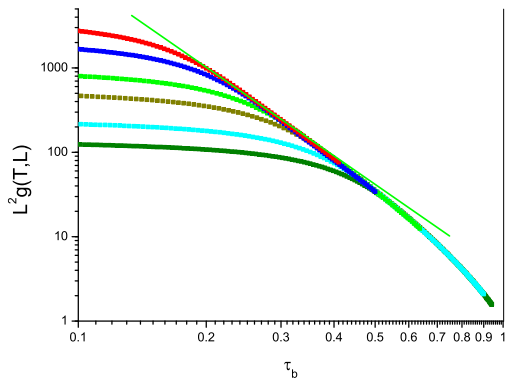}
  \caption{(Color on line) Gaussian 2D ISG. $L^2 g(\tau_{b})$ against
    $\tau_{b}$. Sizes $L= 64$, $48$, $32$, $24$, $16$, $12$ top to
    bottom. Same color coding as in Fig.~\ref{fig3}. $g(T,L)$ is the
    Binder cumulant. Line slope $-3.5$.  }\protect\label{fig13}
\end{figure}

\begin{figure}
  \includegraphics[width=3.0in]{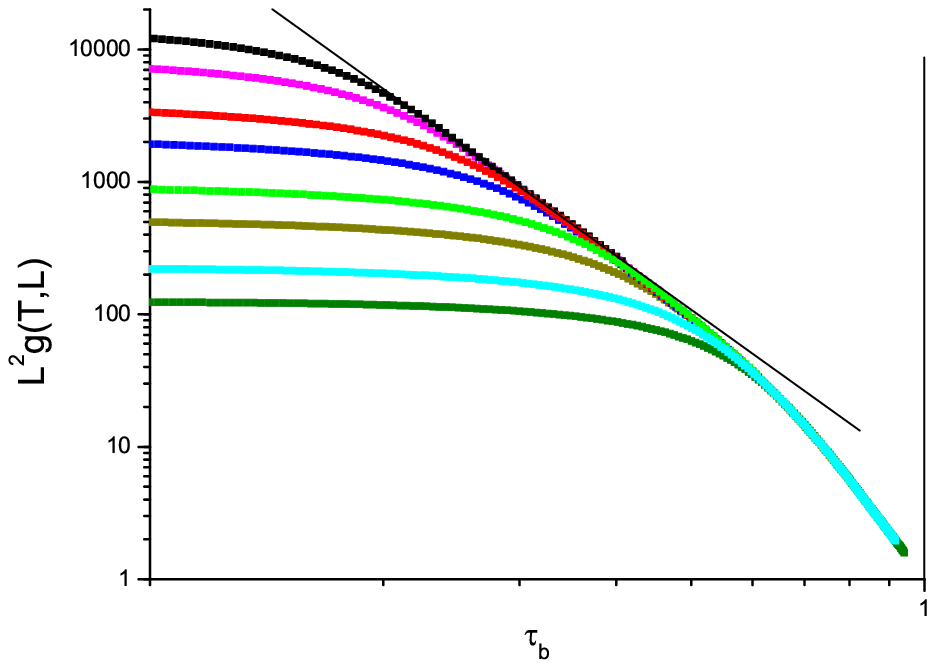}
  \caption{(Color on line) Bimodal 2D ISG. $L^2 g(\tau_{b})$ against
    $\tau_{b}$. Sizes $L= 128$, $96$, $64$, $48$, $32$, $24$, $16$,
    $12$ top to bottom. Same color coding as in
    Fig.~\ref{fig3}. $g(T,L)$ is the Binder cumulant. Line slope
    $-4.2$.  }\protect\label{fig14}
\end{figure}

\section{Appendix I: Binder cumulant}
The ferromagnetic Binder cumulant has been extensively exploited in
the FSS limit regime very close to criticality for its properties as a
dimensionless observable.  In addition its ThL properties can also be
studied.  In Ising ferromagnets, the critical exponent for the second
field derivative of the susceptibility $\chi_{4}$ (also called the
non-linear susceptibility), is \cite{butera:02}
\begin{equation}
  \gamma_{4}=\gamma +2\Delta = \nu d + 2\gamma
\end{equation}
The non-linear susceptibility $\chi_{4}$ is directly related to the
Binder cumulant, \cite{privman:91} Eq.~{10.2}, through
\begin{equation}
  g(\beta,L) = \frac{-\chi_{4}}{L^d\chi^{2}} = \frac{3\langle
    m^2\rangle - \langle m^4\rangle}{\langle m^2\rangle^2}
\end{equation}
As $\chi$ scales with the critical exponent $\gamma$, the normalized
Binder cumulant $L^d g(\beta,L)$ scales with the ThL regime critical
exponent $\partial\ln(L^d g)/\partial\ln\tau = (\nu d + 2\gamma) -
2\gamma = \nu d$.  In any $S=1/2$ Ising system the infinite
temperature (independent spin) limit for the Binder cumulant is
\begin{equation}
  g(\infty,N) \equiv 1/N
\end{equation}
where $N$ is the number of spins; $N=L^d$ for a hypercubic lattice.
So $L^d g(\beta,L)$ has an infinite temperature limit which is
strictly $1$, and a large $L$ critical limit (with corrections as for
the other observables):
\begin{equation}\label{gscaled}
  L^d g(\tau_b,L) \sim \tau_b^{-\nu d}\left(1+ \cdots\right)
\end{equation}

Exactly the same argument can be transposed to ISGs (see
Ref.~\cite{klein:91} for $\chi_{4}$ in ISGs). In the particular case
of a $2$D ISG model with $\tau_{b}$ scaling, the critical value for
the derivative $\partial\ln(L^d g(\tau_b,L))/\partial\ln\tau_{b}$ of
the Binder cumulant ThL data extrapolated to $\tau_{b}=0$ is $2\nu/2 =
\nu$ where $\nu$ is once again the correlation length critical
exponent in the $T$ scaling convention. The Binder cumulant data
plotted in the Eq.~\eqref{gscaled} form are shown for the two models
in Figs.~\ref{fig13} and \ref{fig14}. The ThL envelope curves can be
seen by inspection. The derivatives of these curves have already been
shown in Figs.~\ref{fig9} and \ref{fig10}.

It has been suggested that if two models have the same function when
$y=g(\beta,L)$ is plotted against $x= \xi(\beta,L)/L$, it is a proof
of universality. However, because both $L g^{1/d}(\beta,L)$ and
$\xi(\beta,L)$ are controlled by just the same exponent $\nu$ this is
questionable.

\section{Appendix II}
\begin{figure}
  \includegraphics[width=3.0in]{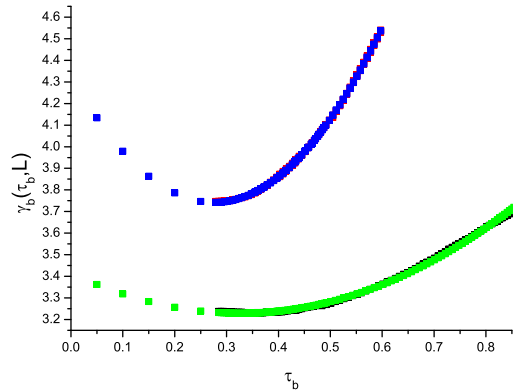}
  \caption{(Color on line) The ThL
    $\partial\ln\chi(\tau_{b})/\partial\ln\tau_{b}$ data for the
    Gaussian model from Fig.~\ref{fig5}, (lower, black) with the
    polynomial fit (green), and for the bimodal model from
    Fig.~\ref{fig6}, (upper, red) with the polynomial fit
    (blue).}\protect\label{fig15}
\end{figure}

\begin{figure}
  \includegraphics[width=3.0in]{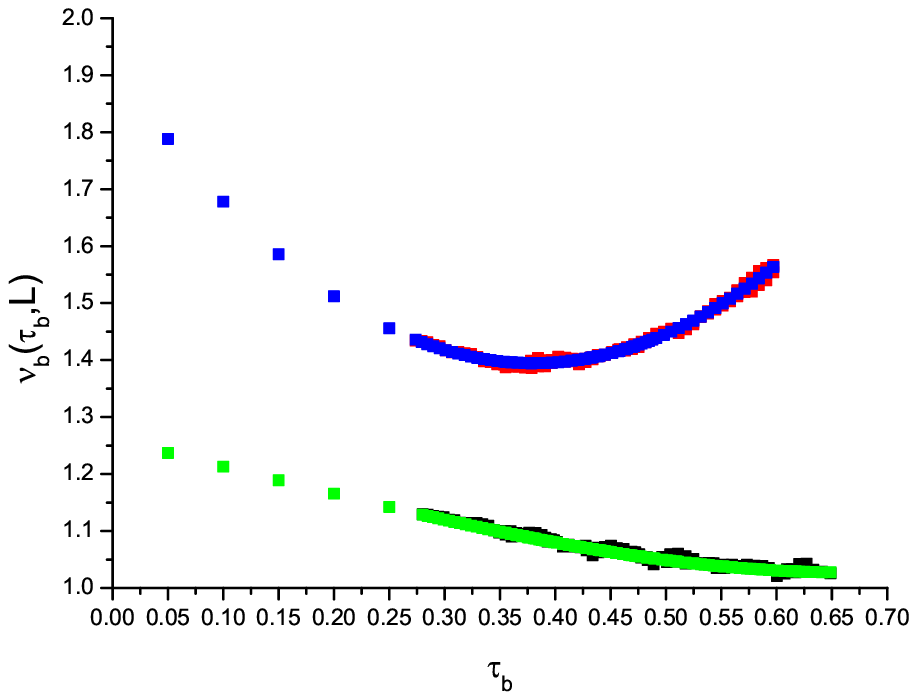}
  \caption{(Color on line) The ThL
    $\partial\ln(T\xi(\tau_{b})/\partial\tau_{b}$ data for the
    Gaussian model from Fig.~\ref{fig7}, (lower, black) with the
    polynomial fit (green), and for the bimodal model from
    Fig.~\ref{fig8}, (upper, red) with the polynomial fit
    (blue).}\protect\label{fig16}
\end{figure}

\begin{figure}
  \includegraphics[width=3.0in]{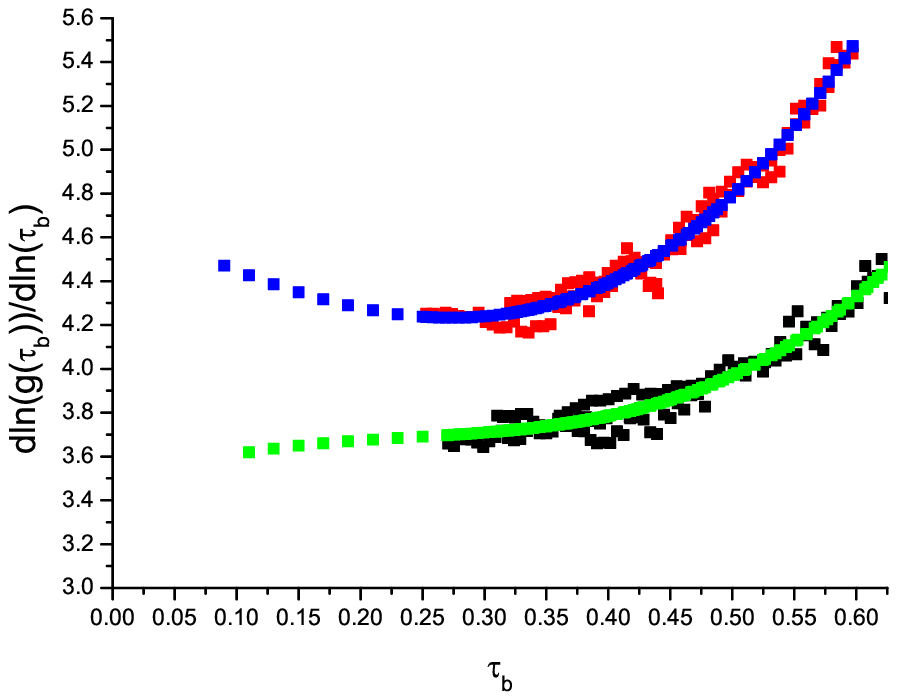}
  \caption{(Color on line) The ThL $\partial\ln
    g(\tau_{b})/\partial\ln(\tau_{b})$ data for the Gaussian model
    from Fig.~\ref{fig9}, (lower, black) with the polynomial fit
    (green), and for the bimodal model from Fig.~\ref{fig10}, (upper,
    red) with the polynomial fit (blue).}\protect\label{fig17}
\end{figure}

\begin{figure}
  \includegraphics[width=3.0in]{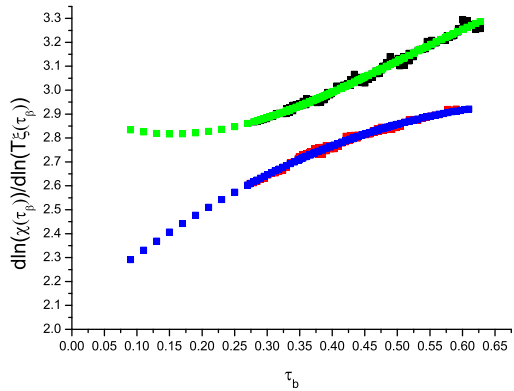}
  \caption{(Color on line) The ThL
    $\partial\ln\chi(\tau_{b})/\partial\ln(T\xi(\tau_{b}))$ data for
    the Gaussian model from Fig.~\ref{fig11}, (upper, black) with the
    polynomial fit (green), and for the bimodal model from
    Fig.~\ref{fig12}, (lower, red) with the polynomial fit
    (blue).}\protect\label{fig18}
\end{figure}

As the data sets do not extend to infinite size, to estimate the
critical $\tau_{b}=0$ limit values from the ThL derivative data in
Figs.~\ref{fig5} to \ref{fig12}, an extrapolation must be made. There
is no definitive method to extrapolate so as to be sure to obtain
exact values of the critical exponents, though data to still larger
sizes would make the task easier.  The most economical choice for
extrapolation is to assume that the ThL derivative data continue to
evolve smoothly and regularly when an extrapolation is made towards
$\tau_{b}=0$ through the smaller $\tau_{b}$ region where no ThL data
are for the moment available. To do this, for each derivative
observable $y(x)$ with $x= \tau_{b}$ we collect together the ThL data
points for all the sizes $L$ up to $x = 0.6$ and make standard
polynomial fits with $3$ or $4$ terms $y(x)= a_{0} + a_{1}x +a_{2}x^2$
or $y(x)= a_{0} + a_{1}x +a_{2}x^2+ a_{3}x^3$ (in fits with larger
numbers of terms the fit parameter values become unstable). Assuming
that each polynomial fit curve extended to $\tau_{b} = 0$ is a good
approximation to the true behavior, each $a_{0}$ is an estimate for
the critical limit value. The $a_{0}$ values for $3$ or $4$ parameter
fits turn out to be similar.  In Figs.~\ref{fig15}, \ref{fig16},
\ref{fig17} and \ref{fig18} the data and fits are shown for
$\partial\ln\chi(\tau_{b})/\partial\ln\tau_{b}$,
$\partial\ln(T\xi(\tau_{b}))/\partial\tau_{b}$ and
$\partial\ln\chi(\tau_{b})/\partial\ln(T\xi(\tau_{b}))$ and
$\partial\ln g(\tau_{b})/\partial\ln\tau_{b}$ for both Gaussian and
bimodal models.  The fits are automatic, so this procedure is
objective and we assume that it is optimal for the available data. All
the Gaussian extrapolated critical values estimated in this way are
close to those expected assuming the published exponents, $\eta = 0$
and $\nu = 3.48(5)$ \cite{hartmann:02}. This implies that the
estimated bimodal critical values are also close to the true critical
limits.

\begin{acknowledgments}
  We are very grateful to Olivier Martin and to Alan Middleton who
  generously allow us access to the specific heat data of
  Ref.~\cite{lukic:04} and of Ref.~\cite{thomas:11} respectively. We
  would like to thank Alex Hartmann for very helpful suggestions.  The
  computations were performed on resources provided by the Swedish
  National Infrastructure for Computing (SNIC) at the High Performance
  Computing Center North (HPC2N) and Chalmers Centre for Computational
  Science and Engineering (C3SE).
\end{acknowledgments}

\end{document}